\documentclass[aps,prd,11pt,onecolumn,nofootinbib,superscriptaddress,amsmath,amssymb]{revtex4-1}

\usepackage{graphicx}
\usepackage[colorlinks=true,linkcolor=blue,citecolor=blue,urlcolor=blue]{hyperref}

\newcommand{\orcid}[1]{\href{https://orcid.org/#1}{\IfFileExists{orcid.png}{\includegraphics[width=10pt]{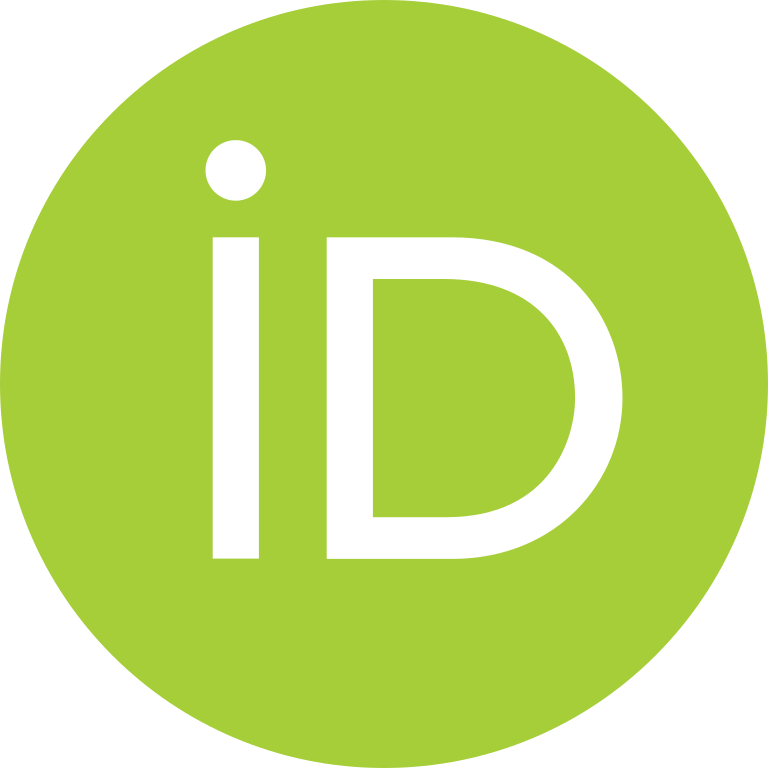}}{\textsuperscript{\scriptsize ORCID}}}}

\begin{document}

\title{Thermal and Casimir Effects of a Scalar Field in a Gravitational-Wave Background}

\author{A. F. Santos \orcid{0000-0002-2505-5273}}
\email{alesandroferreira@fisica.ufmt.br}
\affiliation{Programa de P\'{o}s-Gradua\c{c}\~{a}o em F\'{\i}sica, Instituto de F\'{\i}sica,\\
Universidade Federal de Mato Grosso, Cuiab\'{a}, Brazil}

\author{S. C. Ulhoa \orcid{0000-0001-9994-958X}}
\email{sc.ulhoa@gmail.com}
\affiliation{Instituto de F\'{\i}sica, Universidade de Bras\'{\i}lia, 70910-900, Bras\'{\i}lia, DF, Brazil}
\affiliation{Canadian Quantum Research Center, 106-460 Doyle Ave, Kelowna, British Columbia V1Y 0C2, Canada}

\author{R. G. G. de Amorim \orcid{0000-0001-6532-3087}}
\email{ronniamorim@gmail.com}
\affiliation{Faculdade de Ci\^{e}ncias e Tecnologias em Engenharia, Universidade de Bras\'{\i}lia, 70910-900, Bras\'{\i}lia, DF, Brazil}
\affiliation{Canadian Quantum Research Center, 106-460 Doyle Ave, Kelowna, British Columbia V1Y 0C2, Canada}

\author{F. L. Carneiro \orcid{0000-0001-8329-702X}}
\email{fernandolessa45@gmail.com}
\affiliation{Universidade Federal do Norte do Tocantins, Departamento de F\'{\i}sica,
77824-838 Aragua\'{\i}na, TO, Brazil}

\begin{abstract}
The article deals with the thermal and Casimir properties of a massless
scalar field in a linearized gravitational-wave background. Thermo Field
Dynamics is employed to incorporate temperature and spatial
compactification within a unified real-time framework. Within the local
approximation adopted, the gravitational Stefan--Boltzmann law retains
the usual scalar-field energy density, while the gravitational-wave
polarizations enter the transverse components of the covariant
energy-momentum tensor. Spatial compactification along the direction of
wave propagation produces a negative Casimir energy density and an
attractive longitudinal stress. At finite temperature, the total energy density changes sign at $T_0\simeq 0.329/b$, where $2b$ is the spatial compactification length. This sign change marks a smooth crossover between Casimir-dominated and thermal-dominated regimes.
\end{abstract}

\maketitle

%=========================================================================================
\section{Introduction}
\label{sec:introduction}

Since the advent of general relativity, matter fields have been used to
investigate gravitational dynamics. This approach is rooted in Einstein's
geometric description of gravitation, in which the nonlinear field
equations establish a dynamical interplay between matter and spacetime:
geometry governs the motion of matter, while matter shapes the geometry
in which it evolves, as encapsulated in Wheeler's well-known
formulation~\cite{MTW1973,WheelerFord1998}. Matter fields thus encode
information about the gravitational field, providing a means of
investigating its properties through their behavior. A related
perspective on the role of matter in gravitational physics is provided
by Cooperstock's energy-localization hypothesis, according to which
energy, including gravitational energy, is localized in regions where
the energy-momentum tensor is nonvanishing~\cite{Cooperstock1992}. In
this broader context, scalar fields serve both as models of physical
mechanisms whose underlying nature remains unknown, as in cosmological
scenarios involving
quintessence~\cite{RatraPeebles1988,Wetterich1988,Caldwell1998} and
inflation~\cite{Guth1981,Linde1982,AlbrechtSteinhardt1982}, and as
probes of gravitational backgrounds, as in studies of scalar
perturbations of black holes and their associated quasinormal
modes~\cite{KokkotasSchmidt1999,Berti2009,KonoplyaZhidenko2011}.

The phenomenological role of scalar fields is particularly relevant
here. In the phenomenological approach, an abstract field is introduced
to model an observed phenomenon, and its parameters are constrained by
experimental data~\cite{Copeland2006,Martin2014,Planck2018X}. Such a
description does not specify the underlying physical nature of the
phenomenon but nevertheless constitutes a legitimate approach to
physical inquiry. A similar modeling strategy was widely employed
throughout the nineteenth century in the form of fluid
theories~\cite{Whittaker1951,Harman1982}. Caloric theory provides a
prominent example: the nature of the hypothetical fluid was less
important to the description than its equation of
state~\cite{Fox1971,Brush1976,Chang2003}. Electromagnetic phenomena
were also initially described in terms of fluids~\cite{Heilbron1979}.
The gradual replacement of fluids by fields as the entities employed in
these descriptions did not fundamentally alter the modeling strategy:
an abstract representation was used to formulate quantitative relations
and confront them with
observations~\cite{Hesse1961,Darrigol2000,Maxwell1856}. From this
perspective, scalar fields provide a powerful tool for investigating
gravitational phenomena, particularly when experimental data are
available to constrain their description.

The successful detection of gravitational waves by laser
interferometers provides an important observational setting for such
investigations~\cite{Abbott2016GW150914,Abbott2023GWTC3,SathyaprakashSchutz2009}.
These measurements have enabled studies of fundamental properties of
the gravitational field, as well as of compact objects whose behavior
challenges everyday physical intuition, such as binary neutron-star
systems~\cite{Abbott2017GW170817}. The existence of gravitational waves
remained controversial for many
years~\cite{Kennefick2007,CervantesCota2016}, and methodological
aspects of the interpretation of interferometric data continue to be
critically discussed, as illustrated by the work of Creswell
\textit{et al.}~\cite{Creswell2017}. The development of multimessenger
astronomy, incorporating gravitational waves as an observational
channel, has further expanded the possibilities for exploring the
Universe~\cite{Abbott2017MMA,Meszaros2019}. This broader observational
perspective motivates phenomenological investigations of processes that
are not yet fully understood, including the thermalization of
gravitational
radiation~\cite{Smolin1984,GarfinkleWald1985,PadmanabhanSingh2003,Zhao2009}.

A natural starting point for investigating the thermal properties of
gravitational radiation is the analogy with electromagnetism. The
similarity between Newton's law of gravitation and Coulomb's law
suggests a correspondence between gravitational and electromagnetic
interactions at an appropriate level of approximation, making
electromagnetic phenomena a source of inspiration for gravitational
analogues~\cite{Mashhoon2007,MaartensBassett1998,CostaHerdeiro2008}. In
particular, a Stefan--Boltzmann law for gravitational waves requires a
thermal description of the gravitational
field~\cite{SantosKhanna2016,Ulhoa2019,Ulhoa2021,Santos2021FRW,Santos2021dS}.
The experimentally measured strain of a gravitational wave is a
time-dependent observable, not a measurement of temperature; an
independently measured temperature would therefore provide
complementary information in an analysis involving gravitational
waves. However, there is no consensus on how temperature should be
incorporated into general
relativity~\cite{Tolman1930,TolmanEhrenfest1930,RovelliSmerlak2011,SantiagoVisser2019,Lima2019Tolman,ChuaCallender2025}.
In our view, Thermo Field Dynamics (TFD) provides the most coherent
framework for this purpose. As a real-time formalism, TFD introduces
temperature without replacing the physical time coordinate by an
imaginary-time variable, allowing temporal evolution and thermal
effects to be described within the same
framework~\cite{TakahashiUmezawa1996,Khanna2009}. Formally, TFD is
based on the doubling of the original Fock space through the
introduction of an auxiliary tilde space, together with a Bogoliubov
transformation that constructs a temperature-dependent vacuum.
Expectation values in this thermal vacuum reproduce the corresponding
statistical averages~\cite{TakahashiUmezawa1996,Umezawa1982}. In its
generalized compactification formulation, the formalism incorporates
both thermal periodicity and spatial compactification, providing a
unified treatment of thermal and finite-size effects, including the
Casimir effect at finite temperature~\cite{Khanna2011,Khanna2014}.
Within TFD, the Stefan--Boltzmann law and the Casimir effect of a
scalar field have been obtained in several curved
backgrounds~\cite{Santos2020Schw,Santos2020FRW,SantosKhanna2022}, while
quantum fields and the Casimir effect in gravitational-wave and
weak-field backgrounds have been examined by other
methods~\cite{Gibbons1975,GarrigaVerdaguer1991,Sorge2005,Fulling2007,Nazari2015,Tavakoli2016,Lima2019Casimir,Bezerra2024,deOliveira2026}.
Nazari~\cite{Nazari2015} analyzed vacuum Casimir energies for scalar
and electromagnetic fields between parallel plates in a stationary
weak gravitational background. Tavakoli et al.~\cite{Tavakoli2016}
studied a massless scalar field between plates in the presence of a weak
gravitational wave. In our setup, the scalar field is spatially periodic
along the direction of wave propagation, and we include both vacuum and
thermal contributions.
Taking advantage of the unified description afforded by TFD, our aim in
this work is to investigate the gravitational Stefan--Boltzmann law and
the Casimir effect at both zero and finite temperature.

This article is organized as follows. In Sec.~\ref{sec:scalar_gw}, we
introduce the scalar field coupled to gravity in a linearized
gravitational-wave background and derive the energy-momentum tensor in
the point-splitting formalism. In Sec.~\ref{sec:tfd}, we present the
TFD formalism and its generalized compactification prescription. In
Sec.~\ref{sec:applications}, we apply this framework to investigate the
gravitational Stefan--Boltzmann law and the Casimir effect at zero and
finite temperature. Finally, in Sec.~\ref{sec:conclusions}, we
summarize our main results and present our conclusions.

%=========================================================================================
\section{Scalar field in a linearized gravitational-wave background}
\label{sec:scalar_gw}

We consider a real scalar field $\phi(x)$ propagating in a
four-dimensional curved spacetime described by the metric
$g_{\mu\nu}$~\cite{BirrellDavies1982}. The dynamics of the scalar field
is governed by the action
\begin{equation}
S_\phi =
-\frac{1}{2}\int d^4x\,\sqrt{-g}
\left[
g^{\mu\nu}\partial_\mu\phi\,\partial_\nu\phi
+\left(m^2+\xi R\right)\phi^2
\right],
\label{eq:scalar_action}
\end{equation}
where $g$ denotes the determinant of the metric tensor, $m$ is the
scalar-field mass, $R$ is the Ricci scalar, and $\xi$ represents the
nonminimal coupling between the scalar field and the spacetime
curvature. Throughout this work, natural units $\hbar=c=k_B=1$ are
adopted, together with the metric signature $(-,+,+,+)$ and the
coordinates $x^\mu=(t,x,y,z)$.

The gravitational background is assumed to be described by a weak
gravitational wave propagating along the $z$ direction. Within the
linearized approximation, the spacetime metric can be decomposed as
\begin{equation}
g_{\mu\nu}=\eta_{\mu\nu}+h_{\mu\nu},
\qquad |h_{\mu\nu}|\ll 1,
\label{eq:linear_metric}
\end{equation}
where $\eta_{\mu\nu}=\mathrm{diag}(-1,1,1,1)$ is the Minkowski metric
and $h_{\mu\nu}$ represents a small metric perturbation.

In the transverse-traceless (TT) gauge, the metric perturbation
associated with a plane gravitational wave takes the
form~\cite{MTW1973,SathyaprakashSchutz2009,Maggiore2007}
\begin{equation}
h_{\mu\nu}=
\begin{pmatrix}
0 & 0 & 0 & 0\\
0 & h_+ & h_\times & 0\\
0 & h_\times & -h_+ & 0\\
0 & 0 & 0 & 0
\end{pmatrix},
\label{eq:gw_perturbation}
\end{equation}
where $h_+=h_+(t-z)$ and $h_\times=h_\times(t-z)$ denote the two
independent polarization modes of the gravitational wave. Consequently,
the line element is expressed as
\begin{equation}
ds^2=-dt^2
+\left(1+h_+\right)dx^2
+\left(1-h_+\right)dy^2
+2h_\times\,dx\,dy+dz^2.
\label{eq:gw_line_element}
\end{equation}
Here and throughout the following analysis, terms of second and higher
orders in the metric perturbations are neglected.

The inverse metric, up to first order in the gravitational-wave
amplitudes, is given by
\begin{equation}
g^{\mu\nu}=
\begin{pmatrix}
-1 & 0 & 0 & 0\\
0 & 1-h_+ & -h_\times & 0\\
0 & -h_\times & 1+h_+ & 0\\
0 & 0 & 0 & 1
\end{pmatrix}
+\mathcal{O}(h^2).
\label{eq:inverse_metric}
\end{equation}
Furthermore, owing to the traceless character of the perturbation, the
determinant satisfies
\begin{equation}
\sqrt{-g}=1+\mathcal{O}(h^2).
\label{eq:metric_determinant}
\end{equation}
Within the linearized approximation, the Ricci tensor and the Ricci
scalar associated with the plane gravitational-wave background vanish
to first order in the metric perturbations,
\begin{equation}
R_{\mu\nu}=\mathcal{O}(h^2),
\qquad
R=\mathcal{O}(h^2).
\label{eq:ricci_gw}
\end{equation}
Therefore, the nonminimal curvature-coupling term does not contribute
explicitly to the scalar-field action at linear order.

Substituting Eqs.~(\ref{eq:inverse_metric}) and
(\ref{eq:metric_determinant}) into Eq.~(\ref{eq:scalar_action}), we
obtain
\begin{equation}
\begin{aligned}
S_\phi=\frac{1}{2}\int d^4x\,\Big[
&(\partial_t\phi)^2
-(1-h_+)(\partial_x\phi)^2
-(1+h_+)(\partial_y\phi)^2
+2h_\times\partial_x\phi\,\partial_y\phi\\
&-(\partial_z\phi)^2-m^2\phi^2
\Big]
+\mathcal{O}(h^2).
\end{aligned}
\label{eq:scalar_action_gw}
\end{equation}
This expression explicitly exhibits the coupling between the
scalar-field derivatives and the two gravitational-wave polarization
modes.

In the present investigation, our attention is restricted to a massless
scalar field, $m=0$. The action then reduces to
\begin{equation}
S_\phi=\frac{1}{2}\int d^4x\,\Big[
(\partial_t\phi)^2
-(1-h_+)(\partial_x\phi)^2
-(1+h_+)(\partial_y\phi)^2
+2h_\times\partial_x\phi\,\partial_y\phi
-(\partial_z\phi)^2
\Big]
+\mathcal{O}(h^2).
\label{eq:massless_scalar_action}
\end{equation}

The energy-momentum tensor associated with the scalar field is obtained
by varying the matter action with respect to the inverse metric tensor,
according to
\begin{equation}
T_{\mu\nu}
=-\frac{2}{\sqrt{-g}}
\frac{\delta S_\phi}{\delta g^{\mu\nu}}.
\label{eq:emt_definition}
\end{equation}
The energy-momentum tensor can thus be separated into the Minkowski
contribution and the first-order correction associated with the
gravitational-wave perturbation,
\begin{equation}
T_{\mu\nu}
=
T_{\mu\nu}^{(0)}
+\delta T_{\mu\nu}^{(h)}
+\mathcal{O}(h^2),
\label{eq:emt_decomposition}
\end{equation}
where
\begin{equation}
T_{\mu\nu}^{(0)}
=
\partial_\mu\phi\,\partial_\nu\phi
-\frac{1}{2}\eta_{\mu\nu}
\eta^{\alpha\beta}
\partial_\alpha\phi\,\partial_\beta\phi
\label{eq:emt_flat}
\end{equation}
and
\begin{equation}
\delta T_{\mu\nu}^{(h)}
=
\frac{1}{2}\eta_{\mu\nu}
h^{\alpha\beta}
\partial_\alpha\phi\,\partial_\beta\phi
-\frac{1}{2}h_{\mu\nu}
\eta^{\alpha\beta}
\partial_\alpha\phi\,\partial_\beta\phi.
\label{eq:emt_gw_correction}
\end{equation}
For the TT gravitational-wave background, the energy-momentum tensor
can be expressed as
\begin{equation}
\begin{aligned}
T_{\mu\nu}
={}&
\partial_\mu\phi\,\partial_\nu\phi
-\frac{1}{2}(\eta_{\mu\nu}+h_{\mu\nu})
\Big[
-(\partial_t\phi)^2
+(\partial_x\phi)^2
+(\partial_y\phi)^2
+(\partial_z\phi)^2\\
&-h_+\big(
(\partial_x\phi)^2-(\partial_y\phi)^2
\big)
-2h_\times
\partial_x\phi\,\partial_y\phi
\Big]
+\mathcal{O}(h^2).
\end{aligned}
\label{eq:emt_gw_explicit}
\end{equation}

In the subsequent analysis, the quantum expectation value of this
tensor is evaluated through the point-splitting
formalism~\cite{Christensen1976,DeWitt1975,BirrellDavies1982}, which
allows the energy-momentum tensor to be expressed in terms of two-point
Green functions and provides a suitable framework for investigating the
thermal and finite-size contributions in the presence of a linearized
gravitational-wave background. Within this approach, the products of
quantum fields evaluated at the same spacetime point are replaced by
bilocal expressions involving two distinct points, $x^\mu$ and
$x'^\mu$. Thus, we obtain
\begin{equation}
\begin{aligned}
T_{\mu\nu}(x)
=\lim_{x'\rightarrow x}\tau
\Big\{
&\partial_\mu\partial'_\nu
-\frac{1}{2}(\eta_{\mu\nu}+h_{\mu\nu})
\Big[
-\partial_t\partial'_t
+\partial_x\partial'_x
+\partial_y\partial'_y
+\partial_z\partial'_z\\
&-h_+\left(
\partial_x\partial'_x
-\partial_y\partial'_y
\right)
-h_\times\left(
\partial_x\partial'_y
+\partial_y\partial'_x
\right)
\Big]
\Big\}
\phi(x)\phi(x')
+\mathcal{O}(h^2),
\end{aligned}
\label{eq:emt_bilocal}
\end{equation}
where $\tau$ is the time-ordering operator, and $\partial_\mu$ and
$\partial'_\nu$ denote derivatives with respect to $x^\mu$ and
$x'^\nu$, respectively. The metric perturbations are evaluated at the
point $x$, and the coincidence limit restores the local expression for
the energy-momentum tensor.

In order to investigate the quantum effects associated with the scalar
field, we consider the vacuum expectation value of the energy-momentum
tensor, which provides the relevant physical quantities, such as the
energy density, pressures, and energy fluxes. It is given by
\begin{equation}
\begin{aligned}
\langle T_{\mu\nu}(x)\rangle
=\frac{1}{2}\lim_{x'\rightarrow x}i
\Big\{
&\partial_\mu\partial'_\nu
+\partial_\nu\partial'_\mu
-(\eta_{\mu\nu}+h_{\mu\nu})
\Big[
-\partial_t\partial'_t
+\partial_x\partial'_x
+\partial_y\partial'_y
+\partial_z\partial'_z\\
&-h_+\left(
\partial_x\partial'_x
-\partial_y\partial'_y
\right)
-h_\times\left(
\partial_x\partial'_y
+\partial_y\partial'_x
\right)
\Big]
\Big\}
G_0(x,x')
+\mathcal{O}(h^2),
\end{aligned}
\label{eq:emt_vev_bilocal}
\end{equation}
where
\begin{equation}
iG_0(x,x')
=
\langle 0|\tau[\phi(x)\phi(x')]|0\rangle
\label{eq:green_vacuum}
\end{equation}
is the two-point Green function for the massless scalar field, and
$|0\rangle$ denotes the vacuum state.

Equation~(\ref{eq:emt_vev_bilocal}) establishes a direct connection
between the vacuum expectation value of the energy-momentum tensor and
the two-point Green function of the scalar field. To investigate how
thermal effects and spatial compactification modify the quantum
properties of the system, the TFD formalism is introduced in the next
section. This approach provides a unified framework for incorporating
temperature and finite-size effects through generalized
compactification parameters, allowing the corresponding contributions
to the energy-momentum tensor to be systematically evaluated.

%=========================================================================================
\section{TFD formalism and generalized compactification}
\label{sec:tfd}

TFD is a real-time formalism of quantum field theory at finite
temperature, in which thermal averages are expressed as expectation
values in an enlarged Hilbert
space~\cite{TakahashiUmezawa1996,Umezawa1982,Khanna2009}. The formalism
is based on the doubling of the original degrees of freedom through the
introduction of an auxiliary set of tilde operators associated with the
physical operators. Accordingly, the original Hilbert space
$\mathcal{S}$ is extended to the thermal Hilbert space
$\mathcal{S}_T=\mathcal{S}\otimes\tilde{\mathcal{S}}$, where
$\tilde{\mathcal{S}}$ denotes an auxiliary Hilbert space. This
construction establishes a correspondence between each operator $a$
acting on $\mathcal{S}$ and its tilde counterpart $\tilde{a}$ acting on
$\tilde{\mathcal{S}}$. The relationship between these operators is
governed by the tilde conjugation rules.

Within this framework, thermal and spatial compactification effects are
incorporated through a generalized Bogoliubov transformation,
characterized by a set of compactification parameters denoted by
$\alpha=(\alpha_0,\alpha_1,\alpha_2,\alpha_3)$. These parameters
characterize the compactified spacetime directions and enter the
propagator through the Bogoliubov transformation; in the present
investigation, they are associated with the inverse temperature and a
characteristic spatial compactification length. The TFD formalism
therefore provides a convenient framework for constructing the modified
two-point Green function and determining the corresponding
contributions to the vacuum expectation value of the energy-momentum
tensor.

To establish this formulation, we consider the scalar-field propagator
in the doubled Hilbert space, defined as
\begin{equation}
\begin{aligned}
G_0^{(AB)}(x-x';\alpha)
&=
-i\langle 0(\alpha)|
\tau[\phi^A(x)\phi^B(x')]
|0(\alpha)\rangle\\
&=
\int\frac{d^4k}{(2\pi)^4}\,
e^{-i[k^0(t-t')-\mathbf{k}\cdot(\mathbf{x}-\mathbf{x}')]}
G_0^{(AB)}(k;\alpha),
\end{aligned}
\label{eq:TFD_propagator}
\end{equation}
where $A,B=1,2$ label the physical and auxiliary (tilde) degrees of
freedom, respectively, and $|0(\alpha)\rangle$ represents the
generalized thermal vacuum.

The momentum-space propagator is obtained from the corresponding
zero-temperature propagator through the Bogoliubov transformation
$\mathcal{B}(k;\alpha)$,
\begin{equation}
G_0^{(AB)}(k;\alpha)
=
\left[
\mathcal{B}^{-1}(k;\alpha)
G_0(k)
\mathcal{B}(k;\alpha)
\right]^{AB},
\label{eq:TFD_bogoliubov_propagator}
\end{equation}
where
\begin{equation}
\mathcal{B}(k;\alpha)
=
\begin{pmatrix}
u(k;\alpha) & -v(k;\alpha)\\
\eta v(k;\alpha) & u(k;\alpha)
\end{pmatrix}.
\label{eq:bogoliubov_matrix}
\end{equation}
The functions $u(k;\alpha)$ and $v(k;\alpha)$ satisfy the normalization
condition
\begin{equation}
u^2(k;\alpha)+\eta v^2(k;\alpha)=1,
\label{eq:bogoliubov_normalization}
\end{equation}
with $\eta=+1$ for fermions and $\eta=-1$ for bosons. In the bosonic
case, the function $v^2(k;\alpha)$ is related to the Bose--Einstein
distribution and its generalization to compactified spacetime
dimensions.

The untransformed propagator is represented by the diagonal matrix
\begin{equation}
G_0^{(AB)}(k)
=
\begin{pmatrix}
G_0(k) & 0\\
0 & \eta G_0^*(k)
\end{pmatrix},
\label{eq:TFD_zero_propagator}
\end{equation}
where
\begin{equation}
G_0(k)
=
\frac{1}{(k^0)^2-\lvert\mathbf{k}\rvert^2-m^2+i\epsilon}
\label{eq:scalar_propagator_momentum}
\end{equation}
is the scalar-field propagator in momentum space, and $G_0^*(k)$
denotes its complex conjugate. For the massless scalar field considered
in this work, we set $m=0$.

Using Eqs.~(\ref{eq:TFD_bogoliubov_propagator})--(\ref{eq:TFD_zero_propagator}),
the physical component of the transformed propagator takes the form
\begin{equation}
G_0^{(11)}(k;\alpha)
=
G_0(k)
+\eta v^2(k;\alpha)
\left[
G_0^*(k)-G_0(k)
\right].
\label{eq:TFD_physical_propagator}
\end{equation}
This expression separates the uncompactified propagator from the
contributions induced by the generalized compactification parameters.

The compactification dependence is encoded in the generalized
Bogoliubov coefficient $v^2(k;\alpha)$, which can be expressed
as~\cite{daSilva2002,Khanna2011}
\begin{equation}
v^2(k;\alpha)
=
\sum_{s=1}^{d}
\sum_{\{\sigma_s\}}
2^{s-1}
\sum_{l_{\sigma_1},\ldots,l_{\sigma_s}=1}^{\infty}
(-\eta)^{s+\sum_{r=1}^{s}l_{\sigma_r}}
\exp\left[
-\sum_{j=1}^{s}
\alpha_{\sigma_j}
l_{\sigma_j}
k^{\sigma_j}
\right].
\label{eq:bogoliubov_coefficient}
\end{equation}
Here, $d$ denotes the number of compactified spacetime dimensions,
while $\{\sigma_s\}$ represents the set of all distinct combinations of
$s$ compactified directions. The integers $l_{\sigma_j}$ label the
corresponding compactification modes, and $\alpha_{\sigma_j}$ specifies
the compactification parameter associated with the direction
$\sigma_j$.

For the present investigation, the compactification parameters are
chosen to describe finite-temperature effects and a spatial
compactification along the direction of propagation of the
gravitational wave. This prescription allows the thermal, Casimir, and
mixed contributions to the energy-momentum tensor to be investigated
within a unified framework.

%=========================================================================================
\section{Applications: Thermal and size effects}
\label{sec:applications}

In this section, the thermal and finite-size effects associated with a
massless scalar field in the presence of a linearized
gravitational-wave background are investigated. To this end, three
distinct compactification configurations are considered within the TFD
framework, taking advantage of its topological structure to incorporate
temperature and spatial confinement in a unified description.

The generalized compactification procedure is characterized by the
topology~\cite{Khanna2011,Khanna2014}
\begin{equation}
\Gamma_D^d
=
(\mathbb{S}^1)^d
\times
\mathbb{R}^{D-d},
\qquad
1\leq d\leq D,
\label{eq:compactification_topology}
\end{equation}
where $D$ denotes the spacetime dimension and $d$ represents the number
of compactified dimensions. In this construction, selected directions
of the original spacetime $\mathbb{R}^D$ are compactified into circles
$\mathbb{S}^1$, whose circumferences are specified by the corresponding
compactification lengths. Within the TFD formalism, these
compactifications are encoded in the generalized parameter $\alpha$,
which allows temporal and spatial periodicities to be treated on an
equal footing.

Before considering the different compactification configurations, we
express the vacuum expectation value of the energy-momentum tensor in
the TFD framework. Following the doubling of the Hilbert space and the
corresponding tilde conjugation rules, the energy-momentum tensor is
represented in terms of the doubled field operators. Its expectation
value, given in Eq.~(\ref{eq:emt_vev_bilocal}), can be written as
\begin{equation}
\left\langle T_{\mu\nu}^{(AB)}
(x;\alpha)\right\rangle
=
\lim_{x'\rightarrow x}
\Bigl\{
i\Gamma_{\mu\nu}(x,x')
G_0^{(AB)}(x,x';\alpha)
\Bigr\},
\label{eq:emt_TFD}
\end{equation}
where $A,B=1,2$ denote the doubled degrees of freedom, and
$\Gamma_{\mu\nu}(x,x')$ is the differential operator associated with
the point-split energy-momentum tensor. For the massless scalar field
in the linearized gravitational-wave background, this operator is given
by
\begin{equation}
\begin{aligned}
\Gamma_{\mu\nu}(x,x')
=
\frac{1}{2}
\Big\{
&\partial_\mu\partial'_\nu
+\partial_\nu\partial'_\mu
-(\eta_{\mu\nu}+h_{\mu\nu})
\Big[
-\partial_t\partial'_t
+\partial_x\partial'_x
+\partial_y\partial'_y
+\partial_z\partial'_z\\
&-h_+
\left(
\partial_x\partial'_x
-\partial_y\partial'_y
\right)
-h_\times
\left(
\partial_x\partial'_y
+\partial_y\partial'_x
\right)
\Big]
\Big\}
+\mathcal{O}(h^2).
\end{aligned}
\label{eq:Gamma_TFD}
\end{equation}

To extract the physical contributions induced by temperature and
spatial compactification, we adopt the usual Casimir subtraction
prescription, defined by
\begin{equation}
\mathcal{T}_{\mu\nu}^{(AB)}(x;\alpha)
=
\left\langle
T_{\mu\nu}^{(AB)}(x;\alpha)
\right\rangle
-
\left\langle
T_{\mu\nu}^{(AB)}(x;0)
\right\rangle.
\label{eq:casimir_subtraction}
\end{equation}
Here, the second term is the uncompactified contribution evaluated in
the same gravitational-wave background. This subtraction
removes the common short-distance singularities and isolates the
contributions associated with the compactification parameters.

Using the point-splitting representation, the subtracted
energy-momentum tensor becomes
\begin{equation}
\mathcal{T}_{\mu\nu}^{(AB)}(x;\alpha)
=
\lim_{x'\rightarrow x}
\Bigl\{
i\Gamma_{\mu\nu}(x,x')
\overline{G}_0^{(AB)}(x,x';\alpha)
\Bigr\},
\label{eq:physical_emt_TFD}
\end{equation}
where the subtracted Green function is defined as
\begin{equation}
\overline{G}_0^{(AB)}(x,x';\alpha)
=
G_0^{(AB)}(x,x';\alpha)-
G_0^{(AB)}(x,x';0).
\label{eq:green_subtracted}
\end{equation}
Although the TFD formalism is formulated in the doubled Hilbert space,
physical observables are obtained from the non-tilde sector,
corresponding to $A=B=1$. Therefore, the physical energy-momentum
tensor is expressed as
\begin{equation}
\mathcal{T}_{\mu\nu}(x;\alpha)
=
\lim_{x'\rightarrow x}
\Bigl\{
i\Gamma_{\mu\nu}(x,x')
\overline{G}_0^{(11)}(x,x';\alpha)
\Bigr\},
\label{eq:physical_emt_11}
\end{equation}
where
$\mathcal{T}_{\mu\nu}(x;\alpha)\equiv\mathcal{T}_{\mu\nu}^{(11)}(x;\alpha)$.

Equation~(\ref{eq:physical_emt_11}) establishes the general framework
for evaluating the thermal and finite-size contributions to the
energy-momentum tensor. In the following, three distinct
compactification configurations are investigated in four-dimensional
spacetime. For each case, the generalized Bogoliubov coefficient is
used to construct the corresponding two-point Green function, from
which the energy-momentum tensor is evaluated through the
point-splitting prescription in the linearized gravitational-wave
background.

%-----------------------------------------------------------------------------------------
\subsection{Thermal compactification}
\label{subsec:thermal}

To investigate the effects of finite temperature, we consider the
topology $\Gamma_4^1=\mathbb{S}^1\times\mathbb{R}^3$, characterized by
the compactification parameter $\alpha=(\beta,0,0,0)$, where
$\beta=1/T$ is the inverse temperature.

For this configuration, the generalized Bogoliubov coefficient reduces
to
\begin{equation}
v^2(k^0;\beta)
=
\sum_{l_0=1}^{\infty}
e^{-\beta k^0 l_0}.
\label{eq:thermal_bogoliubov}
\end{equation}
Consequently, the compactification-dependent part of the physical Green
function can be expressed as
\begin{equation}
\overline{G}_0(x-x';\beta)
=
2\sum_{l_0=1}^{\infty}
G_0(x-x'-i\beta l_0 n_0),
\label{eq:thermal_green_sum}
\end{equation}
where $n_0=(1,0,0,0)$ denotes the unit vector along the temporal
direction, and
$\overline{G}_0(x-x';\beta)\equiv\overline{G}_0^{(11)}(x-x';\beta)$
represents the physical component of the subtracted Green function.

Substituting Eq.~(\ref{eq:thermal_green_sum}) into
Eq.~(\ref{eq:physical_emt_11}), the thermal contribution to the
energy-momentum tensor is obtained as
\begin{equation}
\mathcal{T}_{\mu\nu}(x;\beta)
=
2i\lim_{x'\rightarrow x}
\sum_{l_0=1}^{\infty}
\Gamma_{\mu\nu}(x,x')
G_0(x-x'-i\beta l_0 n_0).
\label{eq:thermal_emt}
\end{equation}
This expression provides the starting point for determining the thermal
energy density and the corresponding pressure components in the
linearized gravitational-wave background.

To evaluate the thermal contribution to the energy-momentum tensor, we
first consider the massless scalar-field Green function with the time
displacement,
\begin{equation}
G_0(x-x'-i\beta l_0n_0)
=
-\frac{i}{(2\pi)^2}
\frac{1}{
(x-x'-i\beta l_0n_0)^2
}.
\label{eq:thermal_green}
\end{equation}
Within the local approximation adopted here, the squared interval is
defined by
\begin{equation}
\begin{aligned}
(x-x'-i\beta l_0n_0)^2
={}&
-(t-t'-i\beta l_0)^2
+(1+h_+)(x-x')^2
+(1-h_+)(y-y')^2\\
&+2h_\times(x-x')(y-y')
+(z-z')^2
+\mathcal{O}(h^2).
\end{aligned}
\label{eq:thermal_interval}
\end{equation}

Substituting Eq.~(\ref{eq:thermal_green}) into
Eq.~(\ref{eq:thermal_emt}) and performing the differentiations,
followed by the coincidence limit $x'\rightarrow x$, we obtain the
nonvanishing components of the thermal energy-momentum tensor. The
temporal component is
\begin{equation}
\mathcal{T}_{00}(T)
=
\frac{\pi^2T^4}{30}.
\label{eq:thermal_T00}
\end{equation}
For the transverse components, we find
\begin{equation}
\mathcal{T}_{11}(T)
=
\frac{\pi^2T^4}{90}(1+h_+)
\label{eq:thermal_T11}
\end{equation}
and
\begin{equation}
\mathcal{T}_{22}(T)
=
\frac{\pi^2T^4}{90}(1-h_+).
\label{eq:thermal_T22}
\end{equation}
The longitudinal component is given by
\begin{equation}
\mathcal{T}_{33}(T)
=
\frac{\pi^2T^4}{90}.
\label{eq:thermal_T33}
\end{equation}
Furthermore, the cross polarization produces an off-diagonal component,
\begin{equation}
\mathcal{T}_{12}(T)
=
\mathcal{T}_{21}(T)
=
\frac{\pi^2T^4}{90}h_\times.
\label{eq:thermal_T12}
\end{equation}
Therefore, the thermal energy-momentum tensor can be written in matrix
form as
\begin{equation}
\mathcal{T}_{\mu\nu}(T)
=
\frac{\pi^2T^4}{90}
\begin{pmatrix}
3 & 0 & 0 & 0\\
0 & 1+h_+ & h_\times & 0\\
0 & h_\times & 1-h_+ & 0\\
0 & 0 & 0 & 1
\end{pmatrix}
+\mathcal{O}(h^2).
\label{eq:thermal_emt_matrix}
\end{equation}

From Eq.~(\ref{eq:thermal_emt_matrix}), the thermal energy density is
obtained as
\begin{equation}
\rho(T)
=
\mathcal{T}_{00}(T)
=
\frac{\pi^2}{30}T^4,
\label{eq:thermal_energy_density}
\end{equation}
which reproduces the Stefan--Boltzmann law for a massless real scalar
field. Furthermore, the energy-momentum tensor can be expressed in the
perfect-fluid form
\begin{equation}
\mathcal{T}_{\mu\nu}
=
(\rho+p)u_\mu u_\nu
+p\,g_{\mu\nu},
\label{eq:thermal_perfect_fluid}
\end{equation}
where $u^\mu=(1,0,0,0)$ is the four-velocity of the comoving observer
and $p=\rho/3$ is the radiation pressure.

This result shows that, within the local approximation considered here,
the thermal contribution retains the equation of state of a
relativistic radiation fluid, despite the presence of the
gravitational-wave perturbations. The two polarization modes modify the
transverse components of the covariant energy-momentum tensor, while
the thermal energy density remains unchanged at first order in the
metric perturbations. In particular, the off-diagonal component
$\mathcal{T}_{12}$ arises from the cross polarization $h_\times$ and
reflects the nondiagonal structure of the background metric, rather
than an additional physical shear stress.

%-----------------------------------------------------------------------------------------
\subsection{Spatial compactification}
\label{subsec:spatial}

We now investigate the finite-size effects associated with the spatial
compactification of the scalar field along the direction of propagation
of the gravitational wave. For this purpose, we consider the topology
$\Gamma_4^1=\mathbb{S}^1\times\mathbb{R}^3$, characterized by the
compactification parameter $\alpha=(0,0,0,i2b)$, where $2b$ denotes the
compactification length along the $z$ direction. This configuration
gives rise to a Casimir contribution to the vacuum energy-momentum
tensor~\cite{Casimir1948,BrownMaclay1969,Plunien1986,Bordag2001}.

For this compactification, the generalized Bogoliubov coefficient takes
the form
\begin{equation}
v^2(k^3;b)
=
\sum_{l_3=1}^{\infty}
e^{-i2b k^3l_3}.
\label{eq:casimir_bogoliubov}
\end{equation}
Accordingly, the compactification-dependent part of the physical Green
function becomes
\begin{equation}
\overline{G}_0(x-x';b)
=
2\sum_{l_3=1}^{\infty}
G_0(x-x'-2b l_3n_3),
\label{eq:casimir_green_sum}
\end{equation}
where $n_3=(0,0,0,1)$ is the unit vector along the compactified spatial
direction, and
$\overline{G}_0(x-x';b)\equiv\overline{G}_0^{(11)}(x-x';b)$.

Substituting Eq.~(\ref{eq:casimir_green_sum}) into
Eq.~(\ref{eq:physical_emt_11}), we obtain
\begin{equation}
\mathcal{T}_{\mu\nu}(x;b)
=
2i\lim_{x'\rightarrow x}
\sum_{l_3=1}^{\infty}
\Gamma_{\mu\nu}(x,x')
G_0(x-x'-2b l_3n_3).
\label{eq:casimir_emt}
\end{equation}

To evaluate the Casimir contribution, we consider the massless
scalar-field Green function with a spatial displacement along the
compactified direction,
\begin{equation}
G_0(x-x'-2b l_3n_3)
=
-\frac{i}{(2\pi)^2}
\frac{1}{
(x-x'-2b l_3n_3)^2
}.
\label{eq:casimir_green}
\end{equation}
Within the local approximation considered here, the squared interval is
given by
\begin{equation}
\begin{aligned}
(x-x'-2b l_3n_3)^2
={}&
-(t-t')^2
+(1+h_+)(x-x')^2
+(1-h_+)(y-y')^2\\
&+2h_\times(x-x')(y-y')
+(z-z'-2b l_3)^2
+\mathcal{O}(h^2).
\end{aligned}
\label{eq:casimir_interval}
\end{equation}

Substituting Eq.~(\ref{eq:casimir_green}) into
Eq.~(\ref{eq:casimir_emt}), applying the point-splitting differential
operator, and taking the coincidence limit $x'\rightarrow x$, the
nonvanishing components of the energy-momentum tensor are obtained. The
Casimir energy density is given by
\begin{equation}
\mathcal{T}_{00}(b)
=
-\frac{\pi^2}{1440b^4}.
\label{eq:casimir_T00}
\end{equation}
The transverse components are
\begin{equation}
\mathcal{T}_{11}(b)
=
\frac{\pi^2}{1440b^4}(1+h_+)
\label{eq:casimir_T11}
\end{equation}
and
\begin{equation}
\mathcal{T}_{22}(b)
=
\frac{\pi^2}{1440b^4}(1-h_+).
\label{eq:casimir_T22}
\end{equation}
For the longitudinal component, we find
\begin{equation}
\mathcal{T}_{33}(b)
=
-\frac{\pi^2}{480b^4}.
\label{eq:casimir_T33}
\end{equation}
The cross polarization also contributes to the off-diagonal components,
\begin{equation}
\mathcal{T}_{12}(b)
=
\mathcal{T}_{21}(b)
=
\frac{\pi^2}{1440b^4}h_\times.
\label{eq:casimir_T12}
\end{equation}
Therefore, the resulting energy-momentum tensor can be expressed in
matrix form as
\begin{equation}
\mathcal{T}_{\mu\nu}(b)
=
\frac{\pi^2}{1440b^4}
\begin{pmatrix}
-1 & 0 & 0 & 0\\
0 & 1+h_+ & h_\times & 0\\
0 & h_\times & 1-h_+ & 0\\
0 & 0 & 0 & -3
\end{pmatrix}
+\mathcal{O}(h^2).
\label{eq:casimir_emt_matrix}
\end{equation}

Equation~(\ref{eq:casimir_emt_matrix}) shows that spatial
compactification generates a negative vacuum energy density,
\begin{equation}
\rho_{\mathrm{C}}
=
\mathcal{T}_{00}(b)
=
-\frac{\pi^2}{1440b^4}.
\label{eq:casimir_energy_density}
\end{equation}
The corresponding energy-momentum tensor exhibits an anisotropic
pressure distribution, reflecting the preferred direction introduced by
the spatial compactification. Although the gravitational-wave
polarizations explicitly modify the transverse coordinate components,
these components satisfy
\begin{equation}
\mathcal{T}_{ij}
=
-\rho_{\mathrm{C}}\,g_{ij},
\qquad i,j=1,2.
\label{eq:casimir_transverse_stress}
\end{equation}
Thus, within the local approximation considered here, the
gravitational-wave perturbations affect the transverse coordinate
components of the energy-momentum tensor without introducing an
additional anisotropy within the transverse plane. The negative
longitudinal stress corresponds to an attractive Casimir effect along
the compactified direction.

%-----------------------------------------------------------------------------------------
\subsection{Combined thermal and spatial compactification}
\label{subsec:combined}

To investigate the Casimir effect at finite temperature, we consider
the topology
$\Gamma_4^2=\mathbb{S}^1\times\mathbb{S}^1\times\mathbb{R}^2$,
characterized by the compactification parameter
$\alpha=(\beta,0,0,i2b)$. This configuration combines time
compactification with spatial compactification along the $z$ direction.

The corresponding generalized Bogoliubov coefficient takes the form
\begin{equation}
v^2(k;\beta,b)
=
\sum_{l_0=1}^{\infty}
e^{-\beta k^0l_0}
+
\sum_{l_3=1}^{\infty}
e^{-i2bk^3l_3}
+
2\sum_{l_0,l_3=1}^{\infty}
e^{-\beta k^0l_0-i2bk^3l_3}.
\label{eq:mixed_bogoliubov}
\end{equation}
The first two terms describe the purely thermal and zero-temperature
Casimir contributions, respectively, while the third accounts for their
combined effect.

Accordingly, the mixed contribution to the physical Green function is
given by
\begin{equation}
\overline{G}_0^{\,\mathrm{mix}}
(x-x';\beta,b)
=
4\sum_{l_0,l_3=1}^{\infty}
G_0\left(
x-x'
-i\beta l_0n_0
-2bl_3n_3
\right),
\label{eq:mixed_green_sum}
\end{equation}
where $n_0=(1,0,0,0)$ and $n_3=(0,0,0,1)$ denote the temporal and
spatial compactification directions, respectively.

Substituting Eq.~(\ref{eq:mixed_green_sum}) into
Eq.~(\ref{eq:physical_emt_11}), we obtain
\begin{equation}
\mathcal{T}_{\mu\nu}^{\mathrm{mix}}
(x;\beta,b)
=
4i\lim_{x'\rightarrow x}
\sum_{l_0,l_3=1}^{\infty}
\Gamma_{\mu\nu}(x,x')
G_0\left(
x-x'-i\beta l_0n_0-2bl_3n_3
\right).
\label{eq:mixed_emt}
\end{equation}
To evaluate this expression, we consider the massless scalar-field
Green function with simultaneous imaginary-time and spatial
displacements,
\begin{equation}
G_0\left(
x-x'-i\beta l_0n_0-2bl_3n_3
\right)
=
-\frac{i}{(2\pi)^2}
\frac{1}{
\left(
x-x'-i\beta l_0n_0-2bl_3n_3
\right)^2
}.
\label{eq:mixed_green}
\end{equation}

Substituting Eq.~(\ref{eq:mixed_green}) into Eq.~(\ref{eq:mixed_emt}),
performing the differentiations, and taking the coincidence limit
$x'\rightarrow x$, we obtain the nonvanishing components of the mixed
thermal--Casimir energy-momentum tensor. The temporal component is
given by
\begin{equation}
\mathcal{T}_{00}^{\mathrm{mix}}(\beta,b)
=
\frac{2}{\pi^2}
\sum_{l_0,l_3=1}^{\infty}
\frac{
3\beta^2l_0^2-4b^2l_3^2
}{
\left(\beta^2l_0^2+4b^2l_3^2\right)^3
}.
\label{eq:mixed_T00}
\end{equation}
The transverse components are
\begin{equation}
\mathcal{T}_{11}^{\mathrm{mix}}(\beta,b)
=
\frac{2(1+h_+)}{\pi^2}
\sum_{l_0,l_3=1}^{\infty}
\frac{1}{
\left(\beta^2l_0^2+4b^2l_3^2\right)^2
}
\label{eq:mixed_T11}
\end{equation}
and
\begin{equation}
\mathcal{T}_{22}^{\mathrm{mix}}(\beta,b)
=
\frac{2(1-h_+)}{\pi^2}
\sum_{l_0,l_3=1}^{\infty}
\frac{1}{
\left(\beta^2l_0^2+4b^2l_3^2\right)^2
}.
\label{eq:mixed_T22}
\end{equation}
For the longitudinal component, we find
\begin{equation}
\mathcal{T}_{33}^{\mathrm{mix}}(\beta,b)
=
\frac{2}{\pi^2}
\sum_{l_0,l_3=1}^{\infty}
\frac{
\beta^2l_0^2-12b^2l_3^2
}{
\left(\beta^2l_0^2+4b^2l_3^2\right)^3
}.
\label{eq:mixed_T33}
\end{equation}
Furthermore, the cross polarization contributes to the off-diagonal
components,
\begin{equation}
\mathcal{T}_{12}^{\mathrm{mix}}(\beta,b)
=
\mathcal{T}_{21}^{\mathrm{mix}}(\beta,b)
=
\frac{2h_\times}{\pi^2}
\sum_{l_0,l_3=1}^{\infty}
\frac{1}{
\left(\beta^2l_0^2+4b^2l_3^2\right)^2
}.
\label{eq:mixed_T12}
\end{equation}
These expressions describe the combined contribution of thermal and
spatial compactifications to the energy-momentum tensor in the
linearized gravitational-wave background, within the local
approximation considered here.

The complete energy-momentum tensor associated with simultaneous
temporal and spatial compactifications is then obtained by combining
the thermal, zero-temperature Casimir, and mixed contributions,
\begin{equation}
\mathcal{T}_{\mu\nu}(x;\beta,b)
=
\mathcal{T}_{\mu\nu}(x;\beta)
+
\mathcal{T}_{\mu\nu}(x;b)
+
\mathcal{T}_{\mu\nu}^{\mathrm{mix}}
(x;\beta,b).
\label{eq:total_thermal_casimir_emt}
\end{equation}

To further investigate the combined effects of thermal and spatial
compactifications, we focus on the temporal component of the
energy-momentum tensor, which determines the energy density. Although
the double summation appearing in the mixed contribution can be
partially evaluated analytically, the resulting expressions are rather
involved. Therefore, a numerical analysis is adopted to examine the
dependence of the energy density on temperature and compactification
length.

Combining the purely thermal, zero-temperature Casimir, and mixed
contributions, the complete energy density is expressed as
\begin{equation}
\rho(T,b)
=
\frac{\pi^2}{30}T^4
-\frac{\pi^2}{1440b^4}
+
\frac{2}{\pi^2}
\sum_{l_0,l_3=1}^{\infty}
\frac{
3\beta^2l_0^2-4b^2l_3^2
}{
\left(\beta^2l_0^2+4b^2l_3^2\right)^3
}.
\label{eq:total_energy_density}
\end{equation}
The first term represents the Stefan--Boltzmann contribution, the
second corresponds to the zero-temperature Casimir energy, and the
third accounts for the mixed thermal--Casimir effects.

To express the result in dimensionless form, we introduce the parameter
\begin{equation}
\chi=bT=\frac{b}{\beta},
\label{eq:dimensionless_parameter}
\end{equation}
which characterizes the relative importance of the thermal and spatial
compactification scales. Multiplying
Eq.~(\ref{eq:total_energy_density}) by $b^4$, we obtain
\begin{equation}
b^4\rho(\chi)
=
\frac{\pi^2}{30}\chi^4
-\frac{\pi^2}{1440}
+
\frac{2\chi^4}{\pi^2}
\sum_{l_0,l_3=1}^{\infty}
\frac{
3l_0^2-4\chi^2l_3^2
}{
\left(l_0^2+4\chi^2l_3^2\right)^3
}.
\label{eq:dimensionless_energy_density}
\end{equation}
This representation shows that the dimensionless energy density depends
only on the combination $bT$, allowing the thermal and finite-size
contributions to be investigated within a unified framework.

\begin{figure}[htb]
\centering
\includegraphics[width=0.80\textwidth]{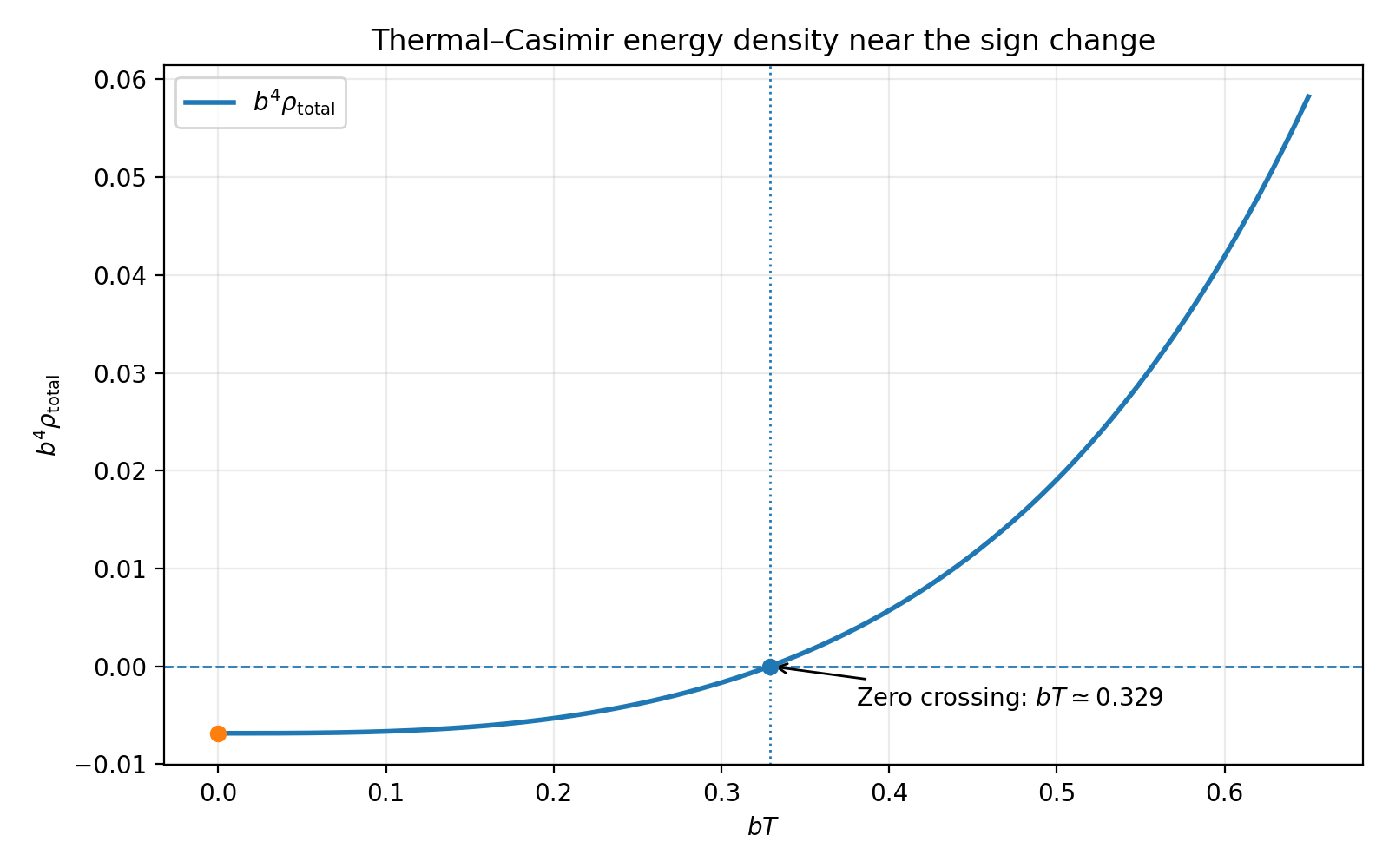}
\caption{Dimensionless energy density $b^4\rho$ as a function of the
parameter $bT$, obtained from the combined thermal and spatial
compactifications. The energy density approaches the negative Casimir
vacuum value $-\pi^2/1440$ in the zero-temperature limit and increases
with temperature, crossing zero at approximately $bT\simeq 0.329$. The
sign change illustrates the competition between the Casimir vacuum
energy and thermal contributions.}
\label{fig:thermal_casimir_density}
\end{figure}

The double summation in Eq.~(\ref{eq:dimensionless_energy_density}) is
evaluated numerically by truncating both summation indices at a
sufficiently large integer $N$. The resulting dimensionless energy
density is displayed in Fig.~\ref{fig:thermal_casimir_density} as a
function of $bT$.

In the zero-temperature limit, the thermal and mixed contributions
vanish, recovering the negative Casimir vacuum energy,
\begin{equation}
\lim_{T\rightarrow 0}b^4\rho(T,b)
=
-\frac{\pi^2}{1440}.
\label{eq:zero_temperature_limit}
\end{equation}
The total energy density vanishes at $bT_0=0.32893$, where the thermal
contribution compensates the negative Casimir vacuum energy. To
characterize the crossing, we calculated the Helmholtz free-energy
density $f$ for the free massless scalar field in flat spacetime at
fixed spatial period $L=2b$ in the limit of infinite transverse area.
The relation $\rho=f-T(\partial f/\partial T)_L$ reproduces
Eq.~(\ref{eq:total_energy_density}). The image series for $f$ converges locally uniformly in a
complex neighborhood of each positive temperature, so $f$ and its
temperature derivatives are analytic at $T_0$. Using the energy density
$\mathcal{T}_{00}$ and longitudinal pressure $\mathcal{T}_{33}$, we
obtained the entropy density $s=(3\mathcal{T}_{00}-\mathcal{T}_{33})/(2T)$
and the heat capacity per unit volume $c_V=T(\partial s/\partial T)_L$.
At the crossing, $b^3s\simeq0.03120$ and $b^3c_V\simeq0.06358$. The
sign change is a smooth thermal--Casimir crossover, with no
thermodynamic phase transition at $T_0$.

%=========================================================================================
\section{Conclusions}
\label{sec:conclusions}

In this work, we investigated the coupling of a gravitational wave to
a massless scalar field to describe thermal and finite-size effects
within TFD. The thermal vacuum construction was used to obtain the
gravitational Stefan--Boltzmann law, while spatial compactification
along the direction of wave propagation allowed us to examine the
Casimir effect at both zero and finite temperature. These
configurations were treated within a unified real-time framework, to
first order in the gravitational-wave amplitudes and within the local
approximation adopted.

The gravitational Stefan--Boltzmann law exhibits the expected
fourth-power dependence on temperature, with the longitudinal pressure
satisfying $\mathcal{T}_{33}=\mathcal{T}_{00}/3$, consistent with the
radiation equation of state. At zero temperature, spatial
compactification produces a negative vacuum energy density and an
attractive longitudinal stress proportional to $b^{-4}$. These results
demonstrate the consistency of the TFD description of the purely
thermal and zero-temperature Casimir configurations.

The total energy density changes sign at $T_0\simeq0.329/b$, marking a
smooth crossover. In the flat periodic system, the free energy is
analytic at $T_0$, which rules out a thermodynamic phase transition at
the crossing. The same zero occurs in the local approximation used for
the gravitational-wave background.

The gravitational-wave polarizations nevertheless enter explicitly into
the transverse covariant components of the energy-momentum tensor. The
plus polarization modifies $\mathcal{T}_{11}$ and $\mathcal{T}_{22}$
with opposite signs, while the cross polarization appears in
$\mathcal{T}_{12}$. Within the approximation considered, these
components are proportional to the transverse metric, so their
coordinate dependence does not by itself establish an additional local
pressure anisotropy. The crossover temperature and the polarization
dependence thus characterize distinct aspects of the result: the former
identifies the thermal change in sign of the energy density, while the
latter describes how the transverse tensor components reflect the
gravitational-wave background.

These results motivate further investigation of whether a measurement
of forces in a specified experimental configuration could provide
information complementary to interferometric strain measurements.
Establishing such a possibility requires relating the calculated
energy-momentum tensor to the response of the measuring apparatus and
distinguishing the wave-dependent contribution from the thermal and
Casimir backgrounds. In this setting, the crossover temperature provides
a reference scale for examining the interplay between thermal,
finite-size, and gravitational-wave effects.

%=========================================================================================
\begin{acknowledgments}
A.~F.~Santos is partially supported by the Brazilian National Council
for Scientific and Technological Development (CNPq), Grant
No.~312406/2023-1. F.~L.~Carneiro acknowledges financial support from
CNPq through a Productivity Fellowship (PQ), Grant No.~303574/2026-7.
\end{acknowledgments}

\section*{Declarations}

\noindent\textbf{Conflict of interest:} The authors declare no conflict
of interest.

%=========================================================================================

\end{document}